# Deep Learning for Automated Experimentation in Scanning Transmission Electron Microscopy

Sergei V. Kalinin,[1] Debangshu Mukherjee,[2] Kevin Roccapriore,[3] Ben Blaiszik,[4,5] Ayana Ghosh,[2] Maxim Ziatdinov,[2,3] A. Al-Najjar,[2] Christina Doty,[6] Sarah Akers,[6] Nageswara S. Rao,[2] Josh Agar,[7] and Steven R. Spurgeon[8,9]

[1] Department of Materials Science and Engineering, University of Tennessee, Knoxville, TN 37831

[2] Computational Sciences & Engineering Division, Oak Ridge National Laboratory, Oak Ridge, TN 37831

[3] Center for Nanophase Materials Sciences, Oak Ridge National Laboratory, Oak Ridge, TN 37831

[4] Argonne National Laboratory, Data Science and Learning Division, Chicago, IL 60439

[5] University of Chicago, Globus, Chicago, IL 60637

[6] National Security Directorate, Pacific Northwest National Laboratory, Richland, WA 99352

[7] Department of Materials Science and Engineering, Drexel University, Philadelphia, PA 19104

[8] Energy and Environment Directorate, Pacific Northwest National Laboratory, Richland, WA 99352

[9] Department of Physics, University of Washington, Seattle, WA 98195

Machine learning (ML) has become critical for post-acquisition data analysis in (scanning) transmission electron microscopy, (S)TEM, imaging and spectroscopy. An emerging trend is the transition to real-time analysis and closed-loop microscope operation. The effective use of ML in electron microscopy now requires the development of strategies for microscopy-centered experiment workflow design and optimization. Here, we discuss the associated challenges with the transition to active ML, including sequential data analysis and out-of-distribution drift effects, the requirements for the edge operation, local and cloud data storage, and theory in the loop operations. Specifically, we discuss the relative contributions of human scientists and ML agents in the ideation, orchestration, and execution of experimental workflows and the need to develop universal hyper languages that can apply across multiple platforms. These considerations will collectively inform the operationalization of ML in next-generation experimentation.

# I. Introduction

The most widely used definition of materials science is the study of structure-property relationships and mechanisms that underpin them[1, 2]. Such relationships extend across multiple length scales – from atomic arrangements and displacements of the order of angstroms to domains and grain structures of tens to hundreds of nanometers to texture relationships across micrometers. The scientific community has expended an enormous amount of scientific effort over the past century to discovering these relationships, finding that most material properties originate from structural phenomena occurring across the length scales from an angstrom to a micron. Whether a defect, a dislocation, a domain boundary, or surface strain, it almost invariably falls between these length scales. However, since optical wavelengths are of the order of a few hundred nanometers, these phenomena exist mostly outside its' purview, although specialized super-resolution fluorescence microscopy has achieved resolution below the Abbe limit down to ~50 nm[3, 4].

Imaging with X-ray photons can reach resolutions of a few tens of nanometers; thus, they are an extraordinarily useful tool in materials science[5-8]. Furthermore, reaching the resolution limits of X-rays can only be done at dedicated synchrotron light sources, which are both expensive and rare. At the same time, these techniques cannot image individual defects, interfaces, dislocations, dopants, or structural information across interfaces – all of which are under the purview of transmission electron microscopy (TEM)[9-12]. TEM is significantly less expensive than a dedicated synchrotron facility, leading to its ubiquity in materials research[11, 13].

Modern electron microscopes are manufactured and operated in two distinct flavors - the first is called conventional TEM (CTEM), where a planar wavefront interacts with the material, and the transmitted wave is recorded with a pixelated electron detector[14]. In the second approach, rather than a plane wave, the electron wavefront is collimated down to an angstrom-scale probe, which is raster scanned across the sample and is thus referred to as scanning TEM (STEM)[15]. Unlike CTEM, the STEM detectors are placed in the diffraction or front focal planes[16]. There is a dizzying variety of detector geometries, but the single most commonly used detector is a ring-shaped single-pixel (integrating) detector, also called an annular detector[17]. With annular detectors, the entire electron count falling on the ring is summed up to generate a single value, and thus a 2D raster scan pattern generates a 2D image from these summed values. In the past decade or so, there has been an interesting divergence - biological EM, especially cryo-EM, has chosen CTEM as their de facto technique, while materials science has increasingly focused on STEM[18, 19]. The reason for this divergence is the development of aberration correction. In STEM, the imaging resolution with annular detectors is governed by the diameter of the electron probe[20-22]. The smaller the probe, the better the resolution. Correction of spherical aberrations can now generate sub-angstrom probes even for accelerating voltages as low as 30kV[23, 24]. Additionally, when STEM detectors are operated in the high-angle annular dark field (HAADF) mode, where the inner collection semi-angle ($\beta$) is larger than ~85 mrads, the obtained image contrast is linear and easily interpretable[17]. Individual atomic columns can be distinguished and differentiated based on this technique.

As a result, STEM has emerged as a powerful tool to study materials structure and functionality across length scales from nanometer to atomic[25] and recently to the deep sub-atomic domains.[26] As atomic-resolution studies of 2D and 3D materials become more routine, these observations have been coupled with computer vision techniques to map minute deviations of atomic positions from high symmetry[12, 27-29]. These approaches have unlocked order parameter fields associated with structural and ferroelectric phase transitions[30-32] and allowed chemical expansion[33] phenomena to be visualized in real space. Such studies, in turn, can be further used to determine the generative physics of materials systems[34-36]. However, a significant drawback of HAADF-STEM is the high dose rates to which samples are exposed: commonly over a million electrons/angstrom$^2$ [37].

One of STEM's most primary advantages is that the electron beam can generate both elastic and inelastic scattering, the latter of which can provide spectral information. In elastic scattering, while the electron energy remains unchanged, the electron beam deviates or deflects from its path due to interaction with the electromagnetic potential in the sample. Coherent elastic scattering is the underlying physical basis for both imaging and diffraction measurements. In contrast, inelastic scattering occurs due to beam electrons exchanging their energy with the material's own orbital electrons and is thus the basis for spectroscopy. The two most widely used spectroscopic techniques in the electron microscope are electron energy loss spectroscopy (EELS) and energy dispersive X-ray spectroscopy (EDX/EDS/XEDS). EELS measurements have provided insights into the single atom level chemistry[38] by probing the low-energy quasiparticles and provided in-depth information on the dark plasmons and light-induced plasmonic phenomena. Apart from resulting in highly interpretable images, another strength of STEM imaging is that when using annular detectors, both the inelastic and elastic scattering information can be captured simultaneously. Modern electron microscopes have thus evolved into complex multimodal equipment that can simultaneously perform imaging, diffraction space experiments, and spectroscopy, leading to the famous observation that it is "*A Synchrotron in a Microscope.*"[20, 39]

Another advancement in the field of electron microscopy has been detector development. As mentioned previously, CTEMs use pixelated detectors, while STEMs conventionally use single-pixel annular detectors. Putting a pixelated detector in place of a single-pixel HAADF detector allows microscopists to use exotic detector geometries in post-processing. Even more exciting is the ability to solve for inverse problems, also called ptychography – where the spatial redundancy in the scanning data is used to solve for the microscope parameters and the sample scattering independent of each other. Yet another area is using such data for nanoscale diffraction experiments to calculate strain or polarization with precision that cannot be matched by HAADF-STEM[40, 41]. The advantages of pixelated detectors have long been known, as Harald Rose speculated on their use cases fifty years back, but the large data volumes they generate have been prohibitive. While standard HAADF-STEM images are on the order of tens of megabytes, these datasets, also called 4D-STEM (2 scanning dimensions, 2 diffraction dimensions in the pixelated detector) are a thousand times larger[42]. It is only recently, that computing power has become widely available to handle them with comparative ease – a prediction that Earl Kirkland accurately made in the late eighties[43].

The developments in STEM imaging and spectroscopic techniques over the last decade have brought forth two closely intertwined problems: namely the analysis of bespoke multidimensional datasets in terms of materials functionality and the development of new approaches toward microscope operation. For the post-acquisition analysis problems, Noel Bonnett envisioned one of the original perspectives on the field early on[44, 45]. In 25 years since his visionary papers were published, many of the predicted advances have indeed been realized. Over the last several years, multiple opinion pieces and roadmaps for ML in (S)TEM have been reported.[46, 47] Particularly over the last four years, the advances in deep learning algorithms and low entry barriers due to the development of high-level languages such as Keras/TensorFlow, JAX and PyTorch have resulted in broad interest in deep learning methods for the tasks such as image segmentation,[48] unsupervised analysis of imaging and spectral data, and learning correlative structure-property relationships.[49]

The rise of large and multimodal electron microscopy datasets has also led to growing use of unsupervised deep learning methods to process such data[50]. There has been however, several recent exciting developments in this field. It was demonstrated last year that spectroscopic and imaging modalities need not be handled separately and can "learn" from each other to generate fused data with SNR levels beyond the reach of any current instrument[51]. Additionally, several papers have recently demonstrated that strain and structural quantification from 4D-STEM datasets can be more effectively performed with pre-trained neural networks on simulated synthetic datasets[52-54]. Such examples have also pushed even CTEM, the old workhorse into new, dose-limited regimes which were not possible earlier. Similar tools now exist even in the synchrotron space, where pre-trained neural networks outperform conventional gradient descent algorithms for ptychography both in speed and data requirements[8, 55].

Despite the progress in instrumentation and post-acquisition data analysis, the basic principles of STEM operation have mostly stayed the same over the last several decades. In almost every case, a human scientist operates the microscope, and it is that scientist who optimizes the microscope performance, explores the sample, and selects regions for further detailed imaging or spectroscopy[56]. The realization of this limitation has led to a strong interest in active learning methods for STEM, in which an ML algorithm controls measurement pathways. Apart from this, simulations can be combined with Bayesian methods to generate idealized experimental routines for data collection, as shown for example in a recent ptychography work[57]. However, despite the strong enthusiasm, there is a growing need for the community to understand what learning can do and what constitutes an automated experiment. In brief, an automated experiment is defined by real-time analysis, which is used to alter the trajectory of experimental decisions. Here we discuss emerging developments in this domain and comment on future opportunities.

**II. ML in microscopy**

Prior to discussing active experiments, we briefly overview several emerging cases for ML in STEM. The vision for the application of ML in microscopy was developed by Bonnett and others almost 25 years ago[45]. However, the practical implementation had to wait for new computational tools and resources to become sufficiently powerful to work with large-dimensional data sets typical for imaging and hyperspectral imaging. The first practical applications of unsupervised ML methods to spectroscopic data dates back to Bosman *et al.* [58].

There are many problems where it is essential to discover the underlying structure-property relationships from multimodal spectroscopic imaging modes in electron microscopy. ML techniques need to disentangle statistical spectroscopic characteristics to facilitate practical interpretations. The challenge with this task is that there is often no good way to form suitable labels. Thus, the philosophy involves learning a constrained low-rank identity function that can disentangle important features of spectroscopic response. Early methods applied linear clustering techniques to associate spectra with characteristic responses[59]. While computationally efficient, these models can only identify groups of characteristic responses and thus cannot deconvolute continuous transformations that occur at interfaces, defects, and topological structures.

Various unsupervised linear ML methods have been applied to accommodate continuous transformations of spectroscopic responses. For example, techniques such as principal component analysis, dictionary learning, and non-negative matrix factorization have been applied across multiple imaging domains.[60, 61] These techniques are deceptively powerful. It is common for these techniques to discover statistical correlations that result in eigenvalues that match physical intuition. However, it is quite rare that the eigenvectors and reconstructions preserve spectroscopic physics. This behavior is the result of several mathematical limitations of the algorithms. First, these methods consider each data point as an independent uncorrelated dimension eliminating important spatial and temporal information. Understanding phase shift or translational shift in diffraction images is out of the question. Secondly, these methods are linear, whereas spectroscopic signals and noise profiles are highly non-linear. These trends must be approximated through overparameterization, or the resulting model is usually significantly underfitting. Due to the computational efficiency and ease of implementation, these methods provide an excellent starting point for machine learning analysis; nonetheless, it is essential that the results are rigorously validated to avoid making incorrect interpretations of underfit data[62, 63].

On the other side of the coin, it is possible to use unsupervised neural networks in the form of autoencoders to disentangle spectroscopic features in hyperspectral images. Autoencoders are a class of neural networks formulated to learn an identity function using three functional blocks. The first block is an encoder that tries to disentangle the statistical information from the raw data; this information is then compressed in an embedding layer in the second block; and the final block is a decoder that takes the compressed information and reconstructs the spectra. Since autoencoders are based on neural networks, it is possible to use logical operations that consider important geometric constraints. For example, convolutions can include translational invariance, harmonic convolutions can learn rotational equivariance, recurrent neurons can consider sequential dependences, and transformers can learn attention maps.

Despite these capabilities, training robust neural networks is challenging, as they are overparameterized. Given enough neurons, autoencoders become universal identity functions capable of memorizing any observable. To make autoencoders practically valuable, it is essential to add bottleneck layers and statistical regularization mechanisms that force the model to learn a compact representation of the data, rather than just memorizing it. In turn, a variety of regularization strategies have been applied. In the first example of a regularized autoencoder for disentangling spectroscopic features in materials spectroscopy, Agar et al. applied scheduled $\|L_1\|$ activity regularization on a ReLu-activated embedding to disentangle sparse, non-negative characteristics of ferroelectric switching in band-excitation piezoresponse force microscopy[64]. In an alternative approach, variational constraints have been included to construct variational autoencoders. Variational autoencoders (VAE) add a Kullback–Leibler (KL)-divergence regularization term to the loss function: $KL[N(\mu_n, \theta_n), N(0,1)]$. This step adds a penalty if the observed distribution in a minibatch deviates from a prior distribution – usually a Gaussian. This process allows learning a smooth latent space that is generally more interpretable.

This concept can be expanded in several ways. For example, a hyperparameter b can be included to modify the KL-divergence regularization term. When regularization scheduling is employed, this term can be used as a disentanglement metric[65]. Such regularization scheduling has been used to disentangle mechanisms of ferroelectric switching[63]. In an alternative modification, learnable geometric transformations can be learned using so-called j(r)-VAEs. These models include a spatial-transforming layer in the embedding to predict an affine transformation grid that learns the geometric orientation and transformation. This information is subsequently passed to the decoder, facilitating the disentanglement of geometric transformations from spectroscopic characteristics.

VAEs have been applied for numerous spectral and image analysis problems emerging in the context of electron and scanning probe microscopy and spectroscopy. These include identification of the elementary building blocks and their distortion in crystalline materials,[66] analysis of rotationally invariant molecular representations in evolving graphene,[67] and semi-supervised learning of molecular shapes from the microscopic data.[66, 68] Here, the repositories such as PyroVED (https://github.com/ziatdinovmax/pyroVED) contain collections of VAE including

rotation- and shift-invariant networks, as well as simple, conditional, semi-supervised, and joint versions.

While these approaches have proven to be quite robust, there are justified concerns about the over reliance on machine learning models to derive physical conclusions. ML methods that include only statistical or only soft physics constraints need to be rigorously validated before any physics conclusions are derived. Conversely, this situation creates tremendous opportunities to include hard physics constraints and governing equations to restrict the output to be physics-conforming, dramatically enhancing interpretability.

In the imaging and spectroscopy domain, much work has been conducted on denoising and sparse reconstructions of images. Controlling spatio-temporal dose distribution to preserve sample integrity is a strong motivator for sparse and/or low dose acquisitions[18]. Some methods require an a priori set of sampling points, determined before an experiment, from which a full reconstruction is later derived. For example, ML based compressive sensing algorithms coupled with microscope hardware allow for collection of a random (or pseudo-random) subset of pixels in an image without sacrificing image quality[16]. Kernel based Gaussian process methods[17] estimate information sharing a priori to constrain and localize variability in high fidelity multivariate reconstructions. These methods use a statically chosen set of sampling points before an experiment is started, but there are existing methods for utilizing on-the-fly information (points updated during experiments) in AE as well, which will be discussed later.

Another broad set of problems involves analyzing images to identify the objects of interest, including localization and characteristics. Many of these problems, including atomic identification, have been extensively explored well before the emergence of deep ML. However, the emergence of the deep DCNNs have made this problem almost routine[48, 69] Note that these semantic segmentation methods are supervised in nature, meaning the algorithm is trained on the human-labelled data and is subsequently applied to a much larger data set.

Visual feedback is a necessary part of image analysis, both when training a ML model and interpreting its results. Values such as accuracy and uncertainty can be calculated to determine the trustworthiness of a model's results, but often the best and most meaningful evaluation comes from seeing the model's results overlayed on top of – or displayed next to the original image. This comparison allows a microscopist to easily see where the model agrees or differs with their expectations, and to interpret and explain results. This type of feedback is also important during the design and training of a ML model. Microscopy images vary significantly between material systems and even within a single material due to noise, imaging artifacts, and beam-induced changes; thus, image acquisition parameters can have a powerful effect on the resulting image. It follows that the preprocessing steps and ML models that produce good results on one image may not translate well to another set of images, making visual feedback throughout the process of designing or tuning a model essential. Some models overcome this barrier and can generalize to multiple material systems by demanding a minimum amount of information about image(s) each time the model is run[20,21]. Providing this information to a model using a command-line interface might not be intuitive to users who would benefit from applying such a model to their own data. A graphical user interface (GUI) designed to guide users through the process of selecting an image, preprocessing it, training the model, and interpreting the results greatly improves the accessibility of ML models to microscopists[70].

The type of user interface needed for a model depends on the intended users. A microscopist may prefer to interface with an existing ML model through a polished GUI such as a web app that guides them smoothly through applying it to their data[70]. One such web app - pyCHIP – allows users to select model parameters that fit their image's feature size by adjusting a slider bar which dynamically updates a grid overlaid on their image. PyCHIP also simplifies the selection of 'support sets' for a few-shot ML model by allowing users to click on regions of the image containing interesting features and adding custom labels[70]. This kind of interface works well for someone looking to apply a model as-is to their dataset. A data scientist on the other hand may appreciate a more dynamic relationship with the model such as through a Jupyter Notebook, allowing them to look under the hood and modify model parameters as needed to tune it to their dataset. The Pycroscopy Python package for example, provides Jupyter notebooks containing well documented code and instructions for utilizing its ML models in different ways[71]. Both web app GUIs and Jupyter Notebook UIs aid in the interpretability and accessibility of ML models for microscopy applications.

The early successes in post-acquisition image and spectroscopy analysis in STEM have naturally generated much excitement in extending these methods towards the real-time data analytics and subsequently automated experiments. In the latter case, the real-time data analysis is used as a basis for a real-time decision making, e.g. choosing the sequence of measurement points for EELS measurements or finetuning the microscope parameters for different parts of the sample.

However, the transition to automated experimentation has proven to be a considerably more complex problem then initially anticipated. First, during an active experiment, the data becomes available sequentially—a very different situation from the static data sets typically used in classical ML methods. Secondly, the microscope performance changes during operation and especially between the subsequent experiments. Consequently, a neural network trained on one day will perform much worse on subsequent days—the celebrated out of distribution drift effect. Third and less recognized until now is the workflow planning process. In other words, a human operator performs the sequence of operations following a specific, implicit, or explicit goal. In certain cases, it is obvious (e.g. when we tune the microscope we aim to improve the resolution), but it is often not trivial to explain a workflow, such as the process of discovery. Hence, a key part of defining automated experiment workflows is the reward, which we next discuss.

**III. Automated experiment: control, algorithm, reward**

Discussion of automated experimentation requires explicit analysis of three dissimilar components. The first and obvious component is the engineering **controls**: the microscope control and command language that determines what operations can be performed and initiated from the external control electronics. The second component is the ML algorithm: that code analyzes the data streaming from the instrument, reduces the data, and generates a sequence of commands returned to the microscope. Third and often overlooked is the **reward**: the perceived goal of the experiment. Given a list of possible operations, an experiment targeting different specific goals will follow different experimental paths. For example, exploring the same sample of the oxide film will lead the scientist to different regions if interested in electrocatalytic activity on the surface vs. magnetoresistive phenomena dominated by the interfaces. Below, we discuss these three elements in more detail.

**III.A. Engineering controls**

Engineering controls currently strongly hinge on the availability of the manufacturer-provided application programming interfaces (APIs) for instrumental control. For example, early work on the electron beam material sculpting[72], direct beam writing[73], atomic manipulation[74], and non-rectangular scans[75] at ORNL relied on the custom modifications of microscope electronics and home built controls. The introduction of the improved version of the NION SWIFT API in the summer of 2021 have permitted unparalleled access to multiple instrument components, including the illumination system, stage, projection system, and detectors and has made these experiments almost straightforward.[76]

Recent approaches based on low-level Python APIs in JEOL microscopes have also achieved task automation for materials focused electron microscopy[77-81]. In the AutoEM system developed by Olszta *et al.*, centralized control is based on asynchronous communication protocols that allow for variable timing between instrument control (action) and instrument readout (reaction), which is essential to account for latencies in both data collection and analysis[79]. Importantly, such latencies can vary depending on the nature of the experiment itself. For instance, in large area statistical montaging, data collection is the bottleneck since it is limited by the speed of mechanical motion (~0.1 fps images), while data analysis via few-shot ML-based classification is fast (~100 fps images). In other cases, such as high-speed in situ experimentation, data collection may fast (1-10k fps images) but a similar data analysis then becomes the bottleneck. In the latter case, only a subset of frames can be processed, limiting the ability to make decisions on rapidly changing phenomena. Here it becomes desirable to have interchangeable models, which a modular system like the AutoEM controller allows. Descriptive models are useful for scenarios where the goal is to identify, describe, and relate features, such as crystal phases, defects, or particles[82, 83].

However, we increasingly require predictive models that can forecast the future state of a chemical or material system for anticipatory control of long-latency instrument parameters. Recurrent neural networks (RNNs) have been successful in forecasting for other domains facing similar problems. Long short-term memory models (LSTMs), a type of RNN, have recently been applied to *in situ* electron energy loss spectroscopy (EELS)[84]. The new model, called EELSTM, inputs a series of prior spectra, such as the initial stages of a chemical reaction, and forecast the future time state of the reaction. They can then be implemented on-the-fly and used to direct an experiment, compensating for long-latency instrument or sample reactions. This area is largely unexplored and presents a tremendous opportunity for more intelligent and impactful automated experimentation.

### III.B. Algorithms

The second key element of automated experimentation is the algorithm, whether machine learning or simple image analysis. Generally, the algorithm has access to the data generated by the microscope at certain level of aggregation, e.g. direct data flow from the detectors or data aggregated at the level of individual images, spectra, or hyperspectral data sets. The algorithms perform certain transformation of the data and return the control signals to the microscope. It is important to note these algorithms can vary in complexity from elementary edge filters or Fourier transforms to complex deep convolutional networks. The choice of the algorithm is determined by the specific task to be accomplished, dimensionality and property of parameter space in which it operates, variability of the data, and by the availability of the compute capabilities and required latencies.

**III.B. Rewards**

Finally, the key and often unrecognized element of the automated experiment is the reward. This is a new element that appears in active learning and more generally stochastic optimization problems, and generally represents the goal of the experiment. Often, the reward function is understood implicitly. For example, tuning the microscope necessitates reducing the effective probe size and minimizing asymmetries. In case when the reward function is well understood and can be readily determined from the data, the problem can be conveniently cast as the optimization over a certain parameter space, for example via Bayesian optimization.[57]

However, reward functions for imaging and spectroscopy experiments can be very non-trivial and are closely aligned with the goal of the experiments. For example, when exploring ferroelectric oxide thin films, the experimentalist can be preponderantly interested in the formation of dislocations and presence of second phase inclusions in the material, properties of the topological defects, evolution of the electronic and phononic properties at structural or topological defects, and many other aspects of material behavior. The sequence of actions taken by the experimentalist will depend on this perceived reward, and the experiment will generally balance the exploration and exploitation component. Here, exploration generally refers to the overview of the structural elements present in the material, whereas exploitation refers to the experimental measurements on specific elements that are a priori known to be of interest. The reward function can change during the experiment. For example, an initial reward function is often calibration and optimization of the imaging conditions, performed using the standard (or well known) sample. Secondly the exploration often aims to ascertain the statistically significant features of the sample and identify outliers (defects, dislocations, etc.). After that, an experimental workflow can be driven by the exploration of objects known to be of interest, or driven by operator curiosity. These considerations necessarily must be considered for the ideation and orchestration of automated experimentation, as will be discussed below.

**IV. On the fly analysis: reconstruction and segmentation**

The first step in automated experimentation is real-time analysis of a data stream from the microscope detectors. This analysis can include denoising and interpolation to improve human operator perception, as well as more complex operations, such as semantic segmentation and feature discovery (e.g. finding atoms on sparse images). This semantically segmented data can inform human operator decisions, or subsequently used to autonomous experiments.

For the pre-acquired data sets, these analyses can be performed via a broad variety of ML algorithms including super resolution networks, compressed sensing, Gaussian processes, or supervised learning. The mathematics behind many of these methods is fairly complex and will not be discussed here. However, the development of open code ecosystem renders them easy to implement once the Python plug-ins for the instrument are available. It is important to note that for any algorithm it is vital to establish the balance between the data and the prior knowledge in some form. For example, compressive sensing and simple Gaussian Processes are purely data driven strategies that seek to interpolate and denoise the data under some general requirements on the smoothness of the image. At the same time, pre-trained neural networks will remember the significant features of their training data, potentially leading to contamination of the processed data, as humorously illustrated in the generative approaches shown in **Figure 1**.

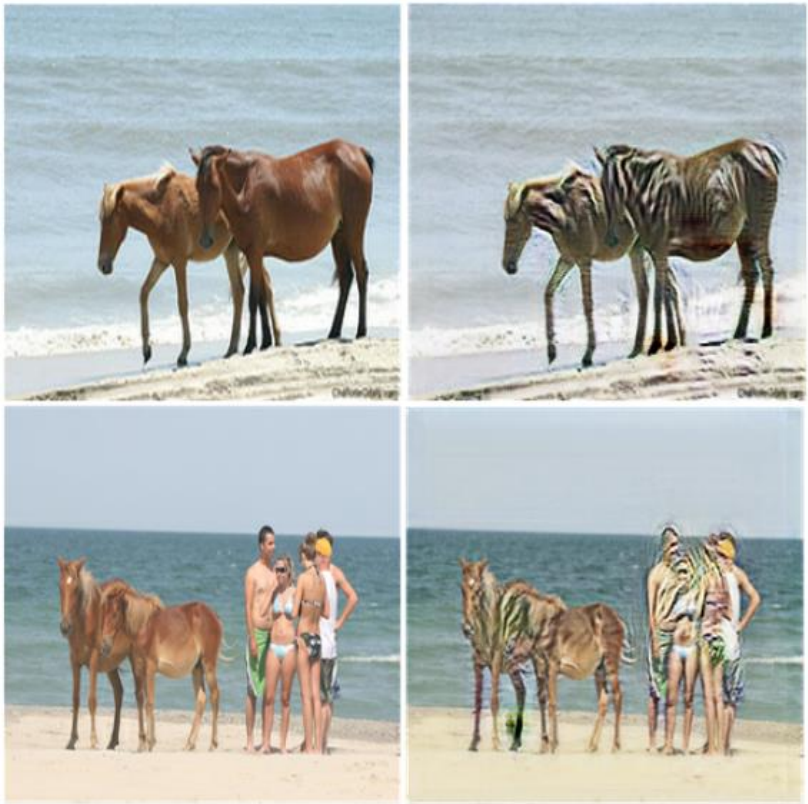

**Figure 1.** Example of the errors introduced via StyleGAN. The generative network allows to transform horse into zebra. However, it also attempts to perform similar operation with humans.

For active learning on microscope these issues are amplified. By definition, information only becomes sequentially available during such an experiment. For unsupervised learning problems, this creates an obvious problem of the analysis of increasing data set at each step, with the interpretation of emerging components changing over acquisition time. For supervised ML problems based on the pretrained model, a key issue becomes out of distribution drift. In other words, the neural network trained for one state of the imaging system will start to fail for different parameter setting.

A typical automated experiment (AE)[85] on the microscope should be capable of performing (a) accurate semantic or chip-level segmentation on images to recognize and extract information on atoms or other relevant features, followed by (b) discovering relationships between beam control parameters and how those affect structure-property relationships of the material under investigation. The final goal is to find out a roadmap, assisted by theory, to obtain target structure by atomic manipulation with the least possible human intervention.

Machine learning/deep learning methods become the obvious choice to accomplish the first task since their applications have proven to be successful at feature detection in many fields such as computer vision, biology, and medicine, as already discussed. These networks tend to work reasonably well on the datasets with large sample sizes, with large variabilities belonging to each category. However, datasets consisting atomically resolved STEM or STM images, the features such as atoms, defects are almost identical (periodic). Changes in imaging conditions on same material system give rise to a different set of data distribution. As a result, one network yielding reasonably accurate prediction may give suboptimal performance on a dataset coming from another set of experimental parameters due to out-of-distribution drift. In addition, deep networks trained on a cumulative dataset with instances from multiple experimental, or simulation conditions are not always sufficient to recognize subtle distinctions in atomic features. Retraining networks by varying hyperparameters every time is also time consuming and limits its applicability within an AE environment. Therefore, to accomplish the task of feature recognition during a live experiment, the network must adapt to changes to the imaging conditions and successfully locate features while quantifying associated uncertainties to make the analyses efficient and accurate.

In our recent work, we have demonstrated how an ensemble learning[86] iterative training (ELIT) workflow[87] can be employed to overcome some of these challenges. Within the ensemble learning (EL) part of the workflow, each new model is trained with different random initialization of weights and random shuffling of training data, with a stochastic weight averaging (SWA) procedure. At the end of training, it averages over multiple points along the trajectory of stochastic gradient descent. The examples on the static datasets on graphene and NiO-LSMO further show how EL is capable to select of artifact-free features and pixel-wise uncertainty maps by combining multiple networks. Switching from image-specific to materials-specific descriptors and training networks on the first frame of images obtained at the initial stages of the experiment can help to account for out-of-distribution effects and improve model performance. Once an initial prediction is obtained, the network is retrained iteratively to realize all features by focusing its attention on features present in the (heavily degenerate) data. It increases the feature detection limit of the network.

Pre-trained ELIT models provide a good start to establish this workflow during a real-time AE. Here, initial training datasets do not necessarily have to come from experiments. They can be obtained by performing first-principles simulations (e.g., density functional theory (DFT), *ab initio* DFT, molecular dynamics (MD)) on the same material system. The models trained on such accurate simulated data may not account for experimental noise or nonidealities. Hence, these generally fail to approximate features from the experimental data. However, the simulated dataset must be carefully augmented to include possible data transformations to minimize the differences between the simulated and experimental structures. Possible augmentations include addition of random zoom-ins, rotations, contrast scaling, and noise. Once the dataset is prepared, ELIT models are trained. During an AE as shown in our recent work,[88] we can utilize the pre-trained ELIT models to detect features from a live data stream by iteratively training the models to analyze local atomic environments, with the possibility of using detector feedback to guide beam conditions such as dwell time.

To successfully assist the next set of experiments and discover structure-property relationships, one must also infuse the fundamental theoretical understanding within the AE loop. The last decades have seen a tremendous growth in the field computational physics, materials science, and related areas, due to significant advancements in computational capabilities including accessible CPU/GPUs, efficient algorithms. Performing physical simulations from the atomistic scale using first-principles techniques to mesoscale/finite scale using quantum Monte Carlo and finite-element methods are all within our reach. As expected, the surge of data that gets generated via these simulations also comes with a wealth of knowledge on thermodynamic, electronic, magnetic properties of materials and physical/chemical phenomenon.

However, a strong disparity still exists between the time, length scale, and associated latencies of running an experiment (e.g., generating one image frame using electron microscope takes fraction of a second) and execution of simulation (e.g., performing geometry optimization on a system in Å to nanometer scales and estimating physical properties take multiple CPU hours). Hence, integration of theory into AE is non-trivial and requires developments of smart approaches[89-92] to enable on-the-fly synchronized learning from the experimental observations and simulations. In our recent work,[93] we have established a workflow that employs deep neural networks to identify atomic features (type and position), followed by constructing simulation objects in an automated fashion to directly pipe those into DFT and MD simulations environment.

Here, various image patches and corresponding simulation objects are chosen by the human expert in the loop to decide the structure fragments from the entire image perspective, suitable to perform simulations. For e.g., segments of graphene that has maintained its ideal coordination after few seconds under the electron beam, may not be the most interesting to investigate compared to regions where it has already begun to form defects. Similarly, it becomes more likely for a successful transition-metal adsorption to take place or dynamic reconstruction of ring arrangements if the regions with symmetry or bond breaking are explored. In addition, performing simulations on the smaller counterparts as compared to the entire large-scale system, also help to match the disparity in the time and length scales, while helping to approximate the overall evolution of the material.

While this workflow is already in place and has led to successful results, it also opens new avenues for exploring structure fragments[94-96] in a broader possible way, which can act as proxies to conduct theoretical investigations during an AE. Moreover, these proxies could be unique in nature that directly relate to the features as recognized by the deep networks to fundamental causal mechanisms. It is established[97-100] how experimental variables can be directly related to properties such as microscopic polarization or lattice parameters. But to determine how these can be controlled, we can incorporate insights from structural distortions connecting to the electronic degrees of freedom, that ultimately control most functionalities[101-104] such as phase transitions, electronic properties (e.g., linear magnetoelectricity, ferromagnetism and polarization and metal-to-insulator transition in transition metal oxides) in solids. Here, these can be exploited as theoretical proxies to establish underlying causal mechanisms[105] to further elucidate the physics involved for assisting next set of measurements.

**V. Direct experiments**

The real time analytics of the STEM data stream further opens the pathway for the implementation of automated experimentation, meaning real-time decision making by the automated agent. During an experiment the human operator typically goes through multiple steps of tuning, selection of region of interest, exploration at different resolution, and spectroscopic measurements. Correspondingly, it makes sense to explore the actions performed by the ML agent in the context of human operation. Following the terminology accepted in active learning community, we refer to the decision-making process adopted by ML agent, i.e. selection of actions based on the state of the system, as ***policy***. The simplest form of an automated experiment is based on predefined policies, which we have previously termed an 'open loop' experiment.[75] In this case, the (pre trained) ML algorithm is used to identify the objects of interest, and predefined sequence of action is taken dependent on observation. Note that in most cases of human-driven experiment the policies are also predetermined (albeit can be very complex), since analysis of the data is usually performed after the experiment and human learning (that will affect policies) typically happens much slower than a single experiment.

A typical example of such experimental measurements uses the real-time analysis of the imaging data stream to identify the a priori known objects of interest, such as specific defect configurations, extended defects, or edges. For example, one of the most common materials science studies is the 'needle in the haystack' search. In this study, it is necessary to be able to both acquire large volumes of representative data—a classic challenge for highly local electron microscopy—and detect features of interest with human-like reasoning. The aforementioned AutoEM system is one example of such a platform that can acquire data repeatably and at scale in an 'open loop' experiment. Data can subsequently be triaged using a few-shot or other ML approach to target desired features of interest. The instrument can then move to those objects, change magnification, select new imaging parameters, etc., constituting a 'closed loop' experiment with rich statistical information, as shown in **Figure 2**. Here the challenge is not simply object detection, but rather the movement of the stage and on-the-fly tuning, which is often quite imprecise and prone to error[54]. Costly and time-consuming registration and image correction routines must be applied to achieve the desired precision for the many hundreds or thousands of acquisitions that might be executed in a typical experiment[106-109]. There is an important opportunity to improve stage hardware, alignment stability, and overall timing to facilitate emerging automated experimentation[78, 110].

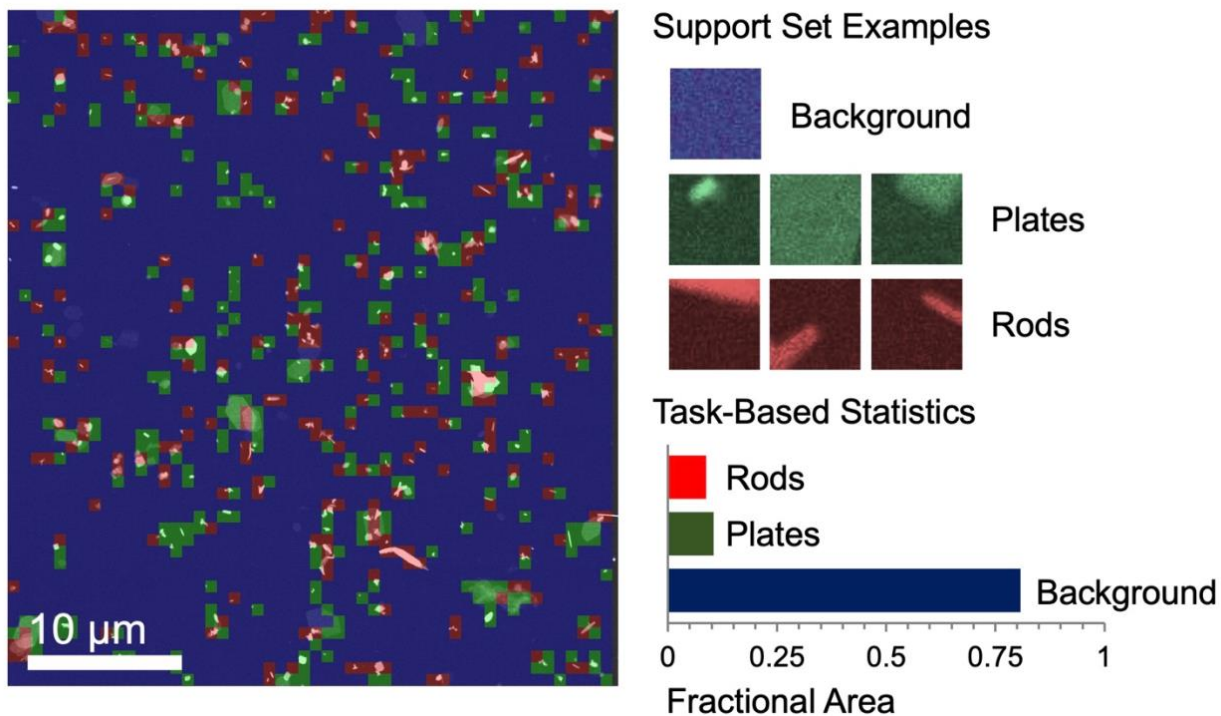

**Figure 2.** Automated task-based statistical analysis of organic photovoltaic precursor synthesis products can be conducted in the STEM, using few-shot analytics combined with low-level instrument control to incorporate human-like decision-making. Adapted from Olszta et al.[75] under CC.-BY-4.0 license.

A similar approach can be applied for the electron-beam modification experiments. In this case, the real time analysis of the STEM data stream can be used to identify objects of interest such as specific atomic configurations. **Figure 3** shows the conversion of live ADF data into classified atomic groups using neural networks within the ensemble learning iterative training (ELIT) scheme,[87] where the computations are performed either on an edge computing device, a connected multi-GPU workstation, or even directly on the instrument machine itself. This was used for realizing controlled experiments where atomic precision of the electron beam relative to specific atomic groups is required – for instance, targeting and ejecting selected S atoms in $MoS_2$.[88] Direct fabrication of atomic devices can be accomplished in this way, as is shown in **Figure 3** with the placement of single (sulfur) vacancy lines in $MoS_2$. We note that in this case the modification policies were also defined prior to the experiment ('open loop').

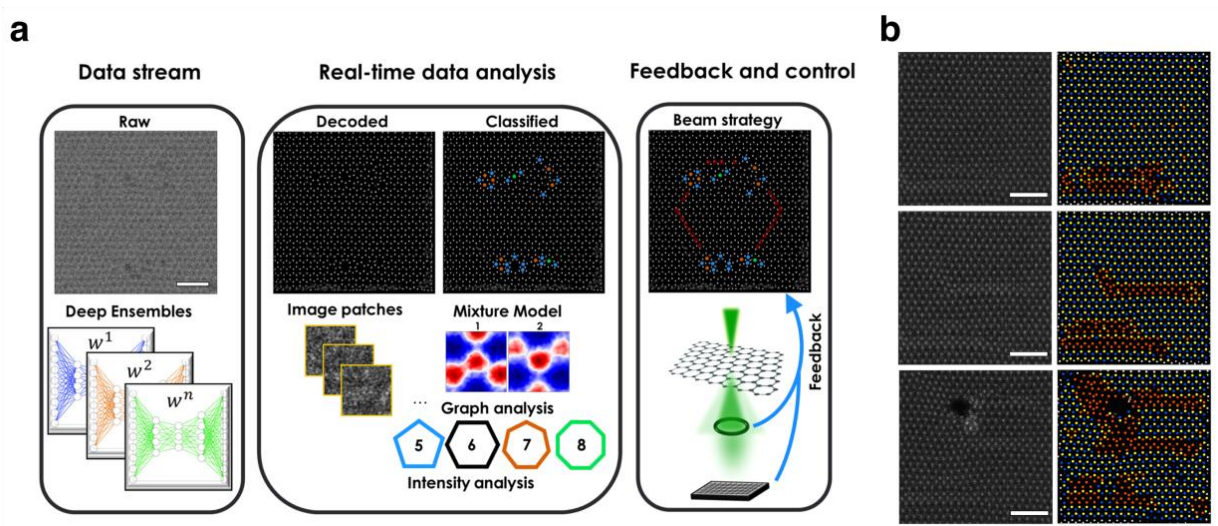

**Figure 3.** Real-time conversion of atomically resolved STEM image data into classified atomic groups for use with atomic fabrication. The neural network-based method utilizes ensemble learning iterative training (ELIT) for robust conversion of STEM image into atomic coordinates, with various methods of classification shown in (a). Classified coordinates are used with control of electron beam position and feedback to target specific atoms, as shown in (b), where left column consists of HAADF-STEM images of $MoS_2$, and right column is the predicted classified atomic groups, where blue represents Mo, yellow represents S, and orange represents single vacancy lines (SVLs). This scheme was used to eject single sulfurs by precisely placing the electron beam on rows of di-sulfur columns and monitoring the ADF feedback to ensure only one sulfur is removed. Reprinted with permission from Ref. [88]

## VI. Indirect experiment and DKL experiments

A more complex task in the context of imaging-spectroscopy is the inverse experiment, where the goal is to discover the structural elements that exhibit a specific spectroscopic feature. We note that in this case the aspect of interest is defined before the experiment (i.e., the policy is fixed), and is assumed to be reduced to one (or several) aspects. For example, for a given EELS spectrum, the aspects of interest can be either peak structure, area under specific peaks, or even pre-trained neural network, as is the case in the EELSTM model.[80] Aspects of interest are typically scalar quantities; therefore, a spectrum may be 'scalarized' into such an aspect by operating on the spectrum in a defined manner. Scalarizing the spectrum in a specific manner allows physics to enter the workflow, which is decided by the user, i.e., the 'human-in-the-loop.' If a spectrum is acquired from within a parent structural image, a relationship can begin to be built between the local structural information from where the spectrum was acquired and an aspect of interest of the spectrum. A similar argument can be made for how a human operator uses their expertise to decide from where they should acquire high-cost measurements, given an observation of a structural space.

These relationships can be discovered autonomously with even only a handful of structure-property pairs on-the-fly and with no prior training, in what is known as deep kernel learning (DKL), and equipped with such structure-property relationships, it allows temporally expensive or high dose measurements to be performed in an intelligent manner. In this way, DKL falls within the category of active learning. This of course has implications for beam-sensitive specimens, but also, it crucially allows exploration of the sample space much more significantly, and the exploration itself is determined by the specified "scalarizer" at the start of the experiment. This means that depending on the scalarizer, different structure-property relationships are learned by the model, and therefore different regions in space will be probed based on that selection. DKL is not limited to EELS,[111] but virtually any multidimensional signal can be used, such as diffraction data in 4D-STEM.[54] How the DKL model learns can be visualized by their exploration pathway (order of acquiring data points). DKL operates by leveraging a deep neural network to embed local structural descriptors (image patches centered at the measurement position) to a low dimensional latent space, where a gaussian process (GP) kernel operates to reconstruct the full data given the acquired data. DKL experiments for both EELS and 4D-STEM (diffraction data), in addition to the impact of choosing different scalarizers, are shown in **Figure 4**.

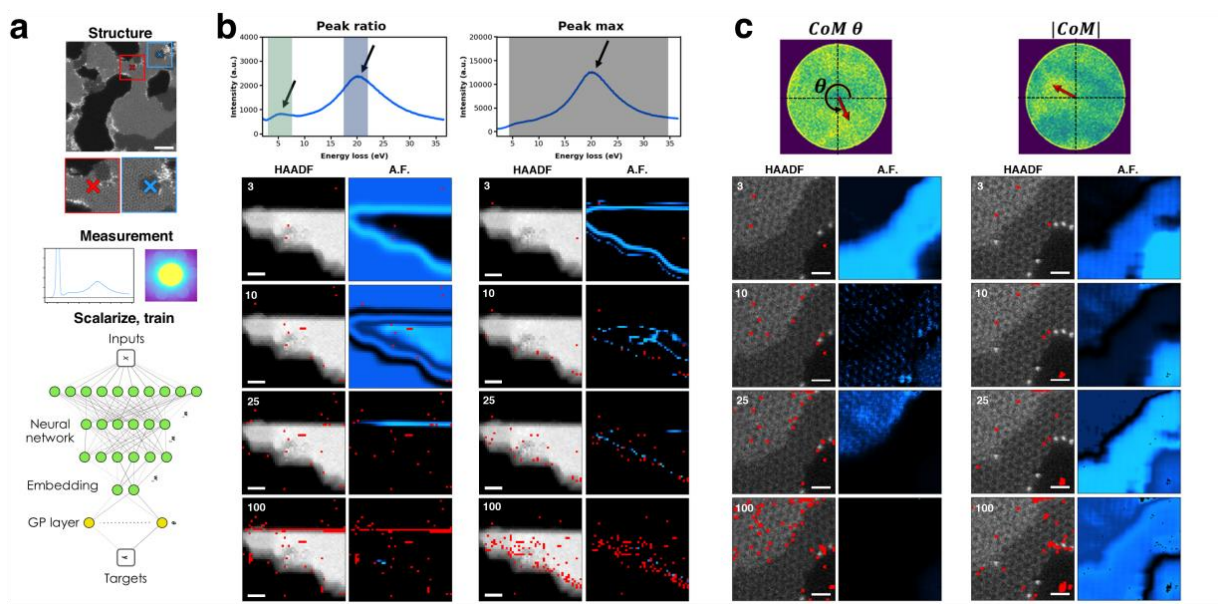

**Figure 4**. DKL in STEM. General schematic workflow depicted in (a) where local image patches are related to an analytical measurement such as a (scalarized) EEL spectrum or diffraction pattern. DKL operating in the EELS domain is shown in (b), where two different scalarizers reveal different exploration pathways (shown as red dots) – peak ratio and peak maximums within specified spectral bands were used to incorporate physics into the model. DKL operating on diffraction data is shown in (c), where the CBED pattern is reduced into the center of mass (CoM), and the magnitude of the CoM and direction of the CoM are used as scalarizers to guide the automated experiment. In both EELS and 4D-STEM versions of DKL, the choice of scalarizer clearly impacts the exploration pathway as well as the structure-property relationship learned. A.F. denotes acquisition function and relates to the learned relationship that combines prediction and uncertainty – the maximum intensity in the A.F. is the next measurement point. Scalebars 50 nm and 1 nm in (b) and (c), respectively. Reprinted (adapted) with permission from [54]. Copyright 2022 American Chemical Society.

For this scheme to be possible on an instrument in real time, there must be access to both the structural image data and the measurement data (EELS, diffraction pattern, etc.). Positioning of the electron beam and readout of the detector signal must also be possible. Note that in many (traditional) systems, the EELS detector operates on a separate machine entirely, making this type of experiment extremely challenging or not possible. The DKL algorithms were implemented on the Nion instruments, where these requirements are easily satisfied *via* access to the scan controller, all detectors, and practically all low-level hardware components through the Nion Swift API.[72] No external hardware is needed, aside from a possible connection to a GPU or multi-GPU workstation for faster model training. While API-level control is not an absolute requirement, it is extremely useful, particularly for complex experiments which are difficult in advance to know what controls are needed. Further, more microscope manufacturers appear to be trending toward increased control of their systems, for example, with JEOL's PyJEM Python package.[75]

**VII. From 2 to many: building characterization workflows**

As discussed above, the operation of the microscope typically involves multiple operator-initiated steps including initial tuning and overview scans and identification of initial objects of interest, as illustrated in **Figure 5**. With these, the imaging, tuning, and spectroscopic cases can be repeated multiple times during the experiment. We note that while this process is familiar to any microscopist, there are several specific aspects that should be enumerated explicitly.

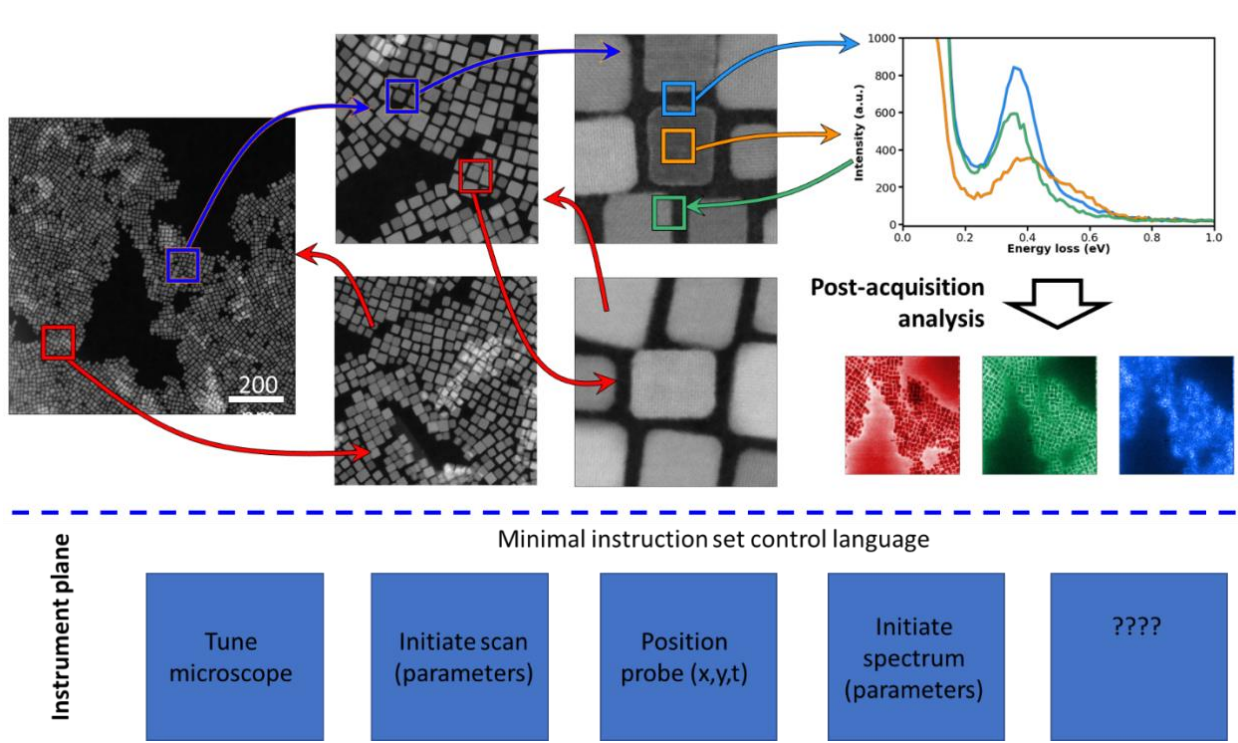

**Figure 5.** Workflow analysis during a STEM-EELS experiment. After the initial tuning, the operator selects the region of interest and performs multiple zoom-in and zoom out stages, retuning, and selection of locations for the spectroscopic studies. This sequence of actions can be represented as a workflow, i.e. sequence of microscope operations performed by the human. The operations implemented by human implicitly define the hyper-language used by the microscope. The workflow design via ML agent is essentially reinforcement learning problem targeting the specific reward via sequence of actions possible within the language.

The first specific aspect is the hyper-language, i.e. the sequence of operations and associated parameters that the human operator can execute. Despite differences among microscopes from different manufacturers and between microscopes of different types (STEM, SEM, SPM), the hyper-languages for many of these systems will be essentially identical and evolve on the convergent paths between different communities. A second key aspect is the workflow, defined here as the sequence of steps performed by the human operator using the available hyper-language. The third component is the reward – meaning the perceived goal of the experiment. Note that while reward is a very familiar concept for scientists in the fields such as stochastic optimization and reinforcement learning, the definition of the reward function in scientific experiment is highly non-trivial and domain dependent. For example, for traditional ML problems such as autonomous driving the reward function is defined as getting from point A to point B in the least time without crashes. Services such as Google Maps provide the preferred path, reducing the problem to driving.

Comparatively, the reward in scientific experimentation is often a very complex function determined by the prior knowledge and perceived goal of the experiment. It can combine the elements of curiosity and general discovery, falsification of hypothesis, or quantitative measurements. For example, it is common to explore new material system in the pure discovery mode, i.e. elucidating the observable structures and morphologies. The more detailed studies can be informed by the prior hypotheses, for example establish the presence of specific structural elements that can account for the macroscopic behaviors. Finally, the experiment can be motivated by the quantitative measurements, e.g. establish the structure of the polarization distribution across the specific ferroelectric domain wall or the strain field around the dislocation.

Following the comparison with the stochastic optimization community, we introduce the concept of policy guiding the experiment. Generally, in reinforcement learning community policy defines the deterministic or probabilistic action taken given the (observed) state of the system. Correspondingly, the policy in the human-driven experiment implicitly defines the selection of the actions by the operator. This process includes elements well established in the RL community, including the exploration and exploitation. The former is exploring the image plane for object of interest. This process can be random, or driven by curiosity, outlier detection, etc. The latter is exploitation, including discovery of specific microstructural elements, hypothesis falsification, or quantitative measurements. While it is difficult to generalize, we argue that the policies used by the human operator will change only weakly during the experiment. For example, depending on the observations the operator can choose a different reward function (i.e. explore the dislocations rather than second phase inclusions), but is unlikely to update the knowledge base during the experiment. These characteristics suggest that workflow planning in the automated experiment can be treated similarly to the reinforcement learning problem, where the reward function is defined prior to the experiment and workflow development includes updating the policies that define the microscope operation possible within the hyper-language commands for a given system.

**VIII.A. Real-time local decision making**

One of the principal challenges with deploying machine learning models is making them practically relevant to the experiment and the scientist. Depending on the experiment, these practical requirements could be vastly different. Thus, it is essential to consider the design of the experiment, algorithms, infrastructure, and hardware when designing real-time analysis for automated experiments in electron microscopy. The primary constraint is that the analysis must happen on a time scale relevant to the experiment. This could mean that the analysis needs to happen in hours to minutes to provide the experimentalist insight to guide the following experiment. If trying to guide active learning and autonomous experiments, the analysis should happen commensurate with the experimental collection speed, usually in minutes to seconds. Finally, it could be required that machine learning be used for real-time controls and triggering that requires millisecond to nanosecond latencies.

Advances in computational resources have made achieving real-time hours-to-minute analysis possible. Excluding fringe cases limited by high-velocity data, many empirical models can be reasonably approximated using neural networks. These processes are simple to implement at scale since they are embarrassingly parallel. Achieving the desired throughput merely requires purchasing and co-locating the correct number of GPUs. When a pre-trained model exists, most electron microscopy experiments can be analyzed in minutes on a local GPU-accelerated workstation.

The implementation of the active learning and automated experiments on STEM necessarily brings forth the consideration of the available computation power, data storage, and data transfer. This can be performed both at the local level and via the integration with distributed cloud resources (and potentially other microscopes). In the former case, data analytics and decision making are performed locally on the computational tools collocated with the microscope. In the second case, the microscope, potentially with some form of point of generation data compression, is integrated with the large-scale cloud resource.

However, achieving real-time analysis can be challenging if a model needs to be trained or fine-tuned for a specific experiment. Model development typically requires hyperparameter tuning to find the optimal parameters to train a model to the specific statistical distributions of the training dataset. Hyperparameter search is generally done using brute force techniques or using e.g., hyperparameter optimization software or architecture tuning tools, each requiring many GPU hours. It is still an open challenge to provide burstable GPU access to experimentalists to train large-scale ML models rapidly. Alternatively, it is possible to reduce the computational complexity of the model and training process. Most machine learning tasks barely consider computational complexity or optimization speed, as providing more computing resources and computational time is comparatively easy. In turn, most machine learning algorithms are overparameterized and over-precise (in terms of bit-depth). The result is flat loss landscapes that are easy to optimize using simple momentum-based stochastic gradient descent optimizers like stochastic gradient descent. Such first-order optimizers sacrifice performance and efficiency for computational simplicity. When optimization speed becomes critical, more advanced optimizers are required. One approach is to use quasi-second-order optimization methods that approximate the Hessian. These methods are inherently more computationally complex but result in significantly better optimization steps, particularly on highly regularized models. Moving beyond first-order optimizers is necessary to achieve real-time electron microscopy analysis.

When trying to automate experiments in minutes to seconds, the primary limitation is established by infrastructural challenges associated with stable and high-speed networking and the high availability of computing resources. The distributed nature of scientific instrumentation makes it uncommon to have networking >1 Gbps at microscopes. Upgrading networking infrastructure requires a reconceptualization of network topologies which tend to bear high costs. In turn, administrators are slow to make high-speed networking (>10 Gbps) available. In an ideal scenario, this limits data velocities to <100 MB/s. In practice, many layers of complexity render connecting scientific instruments to centralized computing impractical. Additional latencies appear from competing network traffic, data ingestion at parallel file systems, and transfer to computing resources. Furthermore, typical job scheduler management systems used in HPC facilities are antithetical to electron microscopy, which requires immediate computational availability when experiments are ready. The solution is to analyze the edge by leveraging technologies such as remote direct memory access; data transferred over networks can be transferred directly to the GPU over the PCIe bus without CPU involvement and thus minimal latency.

Everything becomes more complicated when analyzing real-time controls and triggering that requires inference with < ms-to-nanosecond latencies. General-purpose hardware such as CPUs and GPUs cannot be used as an instruction process, and I/O imparts latency commensurate with scientific experiments. The only viable solution is to directly deploy machine learning models on the extreme edge in programmable logic. The convergence of researchers with knowledge of electron microscopy, machine learning approaches, and programmable logic is currently not well addressed. When deploying machine learning on programmable logic, there are tradeoffs between model performance, computational operations, and hardware resources that must be carefully considered to meet the required figures of merit for an application. This must be achieved through codesign. Recently, several developments have assisted in machine learning codesign on the extreme edge. Packages including Brevitas and QKeras make simpler compressed machine learning models using quantized-aware training. The optimization of these models is assisted by second-order optimizers such as a HAWQ specifically designed for quantization.

Similarly, the design cycles to deploy models on programmable logic (e.g., field-programmable gate arrays [FPGAs]) and application-specific integrated circuits have been facilitated by tools such as HLS4ML and FINN that convert models training in TensorFlow, PyTorch, and Onyx to HLS that can be synthesized into IP-blocks. These frameworks provide a simple abstraction that allows tuning tradeoffs between model complexity, resource utilization, and latency. Developing the physical and cyberinfrastructure to simplify edge machine learning in electron microscopy has tremendous opportunities for growth and could enable entirely new imaging modes and complex atomic-scale manufacturing.

**VIII.B. Microscope as a part of distributed system**

With these considerations in mind, we can design a STEM ecosystem needed to support remote autonomous experiments employing edge computing and storage systems, as shown in **Figure 6**:

- **Local and remote access**: Microscopes are manually operated using custom software installed on control computers co-located with instruments, which are typically connected only to local hub networks. The network access from remote computing nodes is established to both their hardware and software which are protected by access controls and firewalls.

- **OS and software**: Microscope software is typically proprietary and runs on Windows OS, and the measurements are stored on the local computer, and the storage systems, e.g., network attached storage (NAS), often use custom mechanisms and formats. Both control and data access from remote Linux servers may be required, which entails interfacing different OSs, including programming environments, file and data formats, remote file mounts and host firewalls and access mechanisms. For example, in ecosystem shown in Figure 1, files on Windows based NAS are mounted on remote Linux systems using a certain file mounting technique, such as common internet file system (cifs) or secure shell filesystems (sshf) based on NAS system configuration.

- **Networking**: Network connections to control computers carry both measurements and control traffic, typically, over the same IP path and network interfaces. Over long network connections, this non-separation can potentially lead to the loss of instrument control when large measurement transfers occupy the entire available bandwidth. Suitable end-to-end network channels and mechanisms are provisioned between the microscope control computers and remote servers.

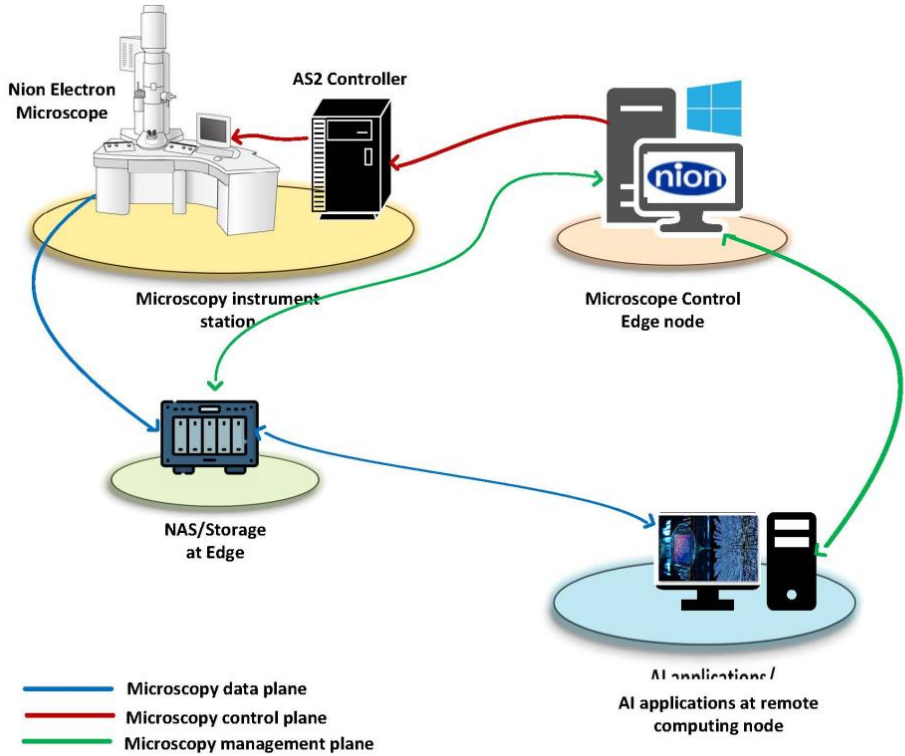

**Figure 6.** The ORNL microscope facility architecture with edge computing and storage system is integrated into networked ecosystem to enable remote, automated experiments orchestrated from remote computing nodes.

An ecosystem design and implementation design are described in Refs. [56, 77, 112] based on separate end-to-end channels for control and data that are serviced by software modules that communicate across Windows and Linux OS. The control channels enable the scientists and automated codes to remotely access the microscope control computer and execute steering commands. The microscope measurements are collected on the NAS and made available on remote computing systems. In addition, a management plane is utilized by the science users and administrators to manage microscope, systems, NAS from the control node.

**Figure 6** shows this design for ORNL ecosystem for the microscope infrastructure. A Nion microscope with an attached AS2 controller is controlled by the instrumentation control computer running Swift software. Upon the completion of a manual or remote command, the measurements from the microscope are transferred and stored on NAS, which is configured to export its file system, thereby making it available for analyses codes that utilize powerful remote computing systems such as servers with multiple GPUs. A control channel is used to remotely access the instrument control computer for sending control commands and parameters to steer the microscope experiments. Upon command execution, the control computer may send back the results or other data. Pyro client-server codes are developed[56, 77] to support remote steering of microscope experiments across the ecosystem. Pyro provides a Python API for network access and is installed it on the control computer's Swift virtual environment and on the remote computing systems. The computing capabilities of remote computing systems, namely, DGX1 with 8 GPUs and Linux server with 2 GPUs, are integrated into this ecosystem, and same approach is applicable to HPC systems which typically are Linux based.

The alternative PNNL ecosystem design is shown in **Figure 7**, utilizing a universal JEOL-based platform running the pyJEM API. The system integrates low-level API commands with an asynchronous, centralized controller that networks all aspects of the instrument and facilitates on-the-fly machine reasoning using sparse data analytics and other ML models. The system abstracts direct control of illumination, stage, and detectors, permitting highly reproducible, standardized experimentation. Importantly, it can rapidly incorporate new analytic models and harness on-premises edge computing hardware, including an NVidia Jetson AGX Xavier-based supercomputing system, for real-time classification, segmentation, and denoising of data. The system is interlinked with PNNL's DataHub lab-level data architecture, permitting automated curation of data and recall for subsequent priming of experimentation. The instrument can be operated in both 'open-loop' and 'closed-loop' modes and observed remotely via a secure terminal connection. A key advantage of this architecture is that the underlying controller and analytics can be deployed on any JEOL hardware system, enabling rich automated experimentation for a broad set of the microscopy community.

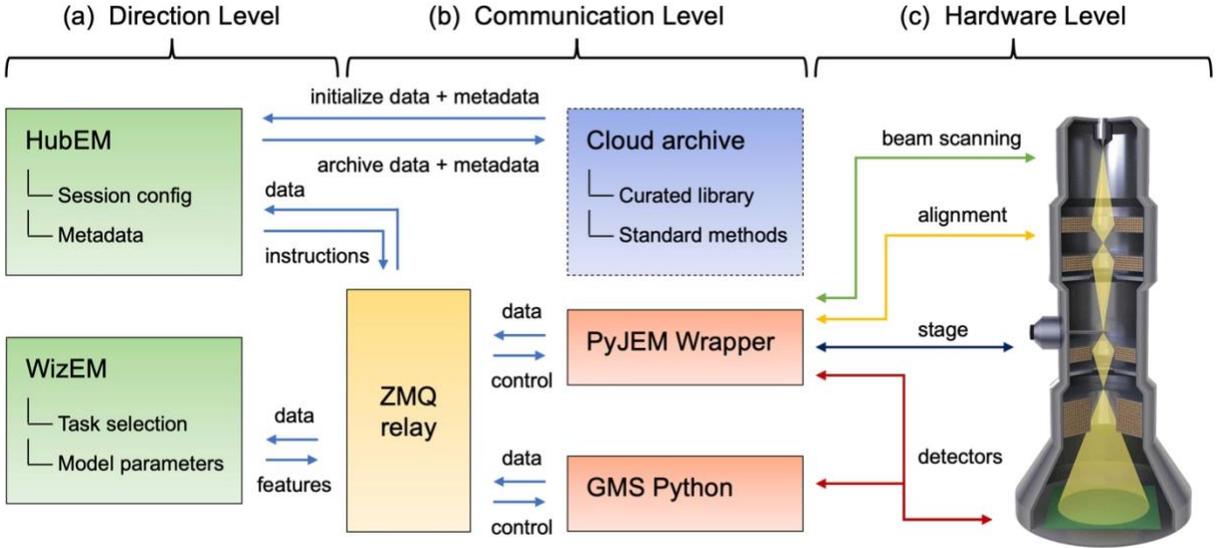

**Figure 7.** The PNNL automated microscopy architecture (AutoEM) is based on asynchronous instrument control of JEOL microscope hardware, leveraging low-level APIs (a), interchangeable sparse data analytics (b), abstracted control (c), and dynamic lab-level data curation for both open- and closed-loop experimentation. Reproduced from Olszta et al.[75] under CC.-BY-4.0 license.

**VIII.C. Data Infrastructure for Shared Instrument Cloud and Automated Microscopy**

New powerful experimental microscopy instruments that are capable of generating 100s of TB/day are being built. Simultaneously, a new generation of powerful computational resources including e.g., Exascale systems at DOE Leadership Computing Facilities and on-demand cloud computing resources (e.g., Amazon Web Services, Microsoft Azure, Google Cloud Compute, NSF Jetstream) while new automation software and services are available to researchers. Thus, there exists an opportunity to weave these varied experimental and computational resources into a cohesive distributed data infrastructure composed of software, services, and resources that facilitate data transfer and movement, advanced computing, flexible policy definition and execution, reusable and shared data, models, and workflows. With such an infrastructure, it will be possible to realize ML-guided real-time analysis and feedback-based microscope operation and to build a connected cloud of instruments vastly increasing the efficiency of instruments and potentially enabling new modes of discovery. We share here key aspects of the enabling distributed data infrastructure followed by an aspirational example of an ML-guided near real-time STEM experiment.

We envision the development of a shared data infrastructure to allow scientists at any networked facility to share data, workflows, and models, instantly accessing the best performing AI algorithms for image analysis from the model repositories and reusing data collected from many instruments from data repositories to enabling an optimized research process. Here, we discuss some of the key data infrastructure components of the instrument cloud.

- **Workflow Repository and Orchestration:** A set of instrument cloud capabilities are abilities to define, orchestrate, share, and reuse workflows. For example, researchers may define workflows, comprised of steps, that coordinate and orchestrate simple and complex actions - perhaps conditional on the results of previous steps or other state. Such actions may include: 1) specifying where and when data should be replicated to locations where processing occurs; 2) defining which functions are applied to data for processing; 3) depositing data and metadata in a shared data repository at the end of the experiment; 4) triggering actions conditional on workflow state, and much more (e.g. Exaworks/Automate/Gladier)
- **Data Collection and Movement:** A core component needed to realize a distributed data infrastructure is the software-defined movement of data between heterogenous systems. These software and services should allow data to be moved, as needed, reliably and seamlessly between e.g., the systems where data are collected, cloud storage, and eventually repositories where the data are made available to the community (e.g. Globus)
- **Flexible Computing:** Microscopy workloads may necessitate computation on a continuum of systems ranging from laptops, cloud, GPU clusters, all the way to Exascale system. Microscopy workflows may require regular retraining of ML models, on-demand model inference, data cleaning and processing, and more (e.g. Parsl, FuncX)
- **Data Repository:** The instrument cloud should have access to repositories of previously collected data, potentially shared across many institutions. These repositories should enable search for existing data, or read and write of data and metadata with defined data structures and metadata schemas. The repository acts to provide 1) raw or processed data for the training of models to guide future experiments, 2) metadata to allow users or agents to search and discover what data are already available and what experiments have been conducted, and 3) a place to share data with the wider community (e.g. MDF/Foundry)
- **Model Repository:** The instrument cloud should also have access to similarly read, write, and search for and shared models. For example a model repository could contain 1) the latest and versions of models trained and used as part to track provenance; and 2) provide to simplify inference with registered models.

Finally, we define an aspirational ML-guided automated STEM experiment flow that leverages the instrument cloud capabilities described above. In this scenario, shown in **Figure 8**, data are collected at a high-powered microscope at ORNL or PNNL that generates data at rates too fast to process with local compute resources. Instead, the data are moved to a processing facility, (e.g., at the nearby Oak Ridge Leadership Computing Facility (OLCF) or PNNL's Constance / Deception Supercomputer). Upon data generation, a script automatically begins streaming the data to the OLCF system for processing. Processed data and metadata are added to a central repository that may also connect data from external resources. At the same time, a policy is defined that automatically initiates the GPU-heavy workload of model retraining upon collection of a new dataset for a specific material. The new model(s) are stored in a model repository that may also access external models. Finally, the trained model and weights are sent back to the data collection for inference using a local GPU. Critical to this flow is seamless and automated data and metadata collection, data registration in a repository, flexible computing, a policy engine to define where and when various compute happens. Similar examples have been developed for common workflows at beamlines at the Advanced Photon Source for processing at the Argonne Leadership Computing Facility for near real-time feedback (e.g. Gladier).

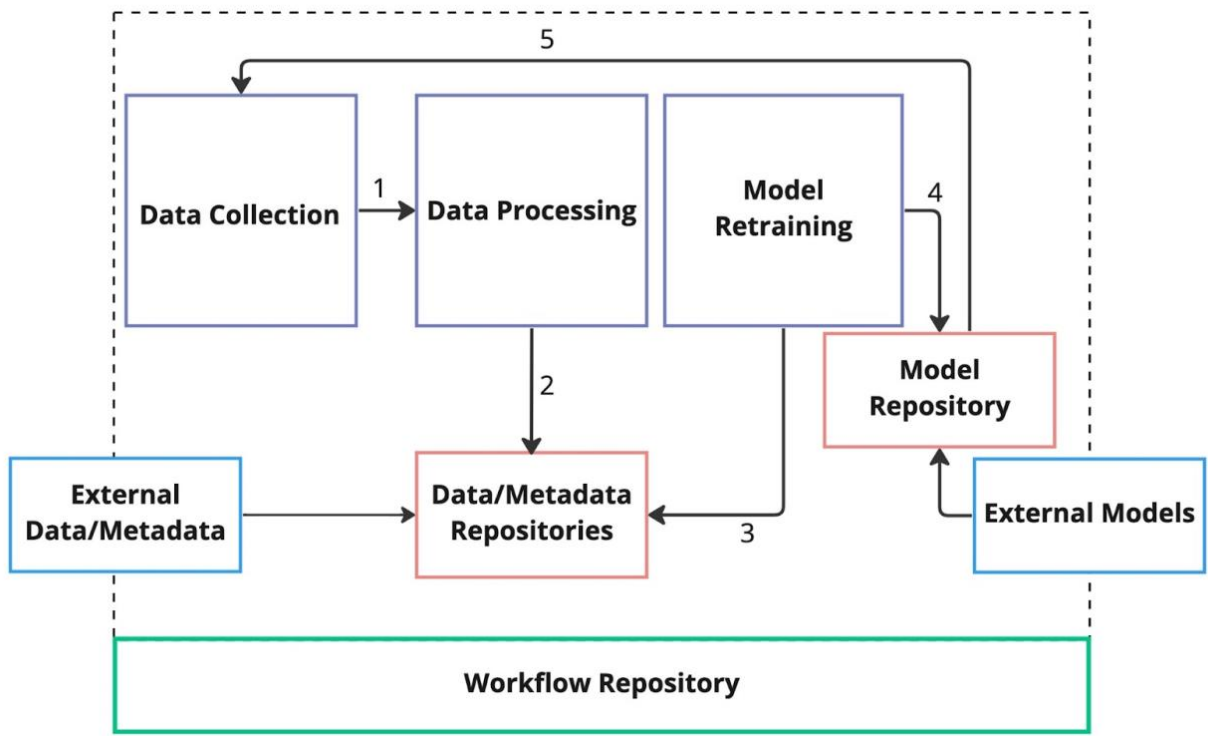

**Figure 8:** An example workflow built using distributed data infrastructure to enable an AI-guided microscopy experiment. This general flow encompasses (1) automated data collection at the microscope, and movement of these data to a processing location; (2) registration of these data in a data repository; (3) periodic or policy-driven retraining of models on data collected in the data repository; (4) registration and versioning of the newest model in a model repository and (5) movement of the model back to the data collection location or inference from a cloud service. External data, metadata and models allow for sharing across workflows.

## IX. Toward the Future

Scanning transmission electron microscopy and spectroscopy has become one of the foundational tools in modern materials science, condensed matter physics, chemistry, and biology. The impact of this technique is directly related to the amounts of quantifiable information on materials structure and properties it can derive. The success of fields such as Cryo EM and small crystal electron crystallography suggest that the availability of the data analysis methods and operational workflows greatly amplifies the value derived from technique developments and suggests tremendous potential for the field growth.

One of the rapidly emerging trends in STEM is the development of the automated experiments. Here, we overview some of the challenges that transition from human-driven to automated experiment EM will bring. On the instrumental side, this necessitates the development of the instrument-level hyper-languages that allow to represent the human operations via minimal primitives. On the ML side, it requires development of the supervised ML algorithms that are stable with respect to the out of distribution drift effects and active learning methods that can be trained on small volumes of data. On the computational and network side, it requires development of edge computing infrastructure capable of supporting rapid analysis and decision making, and connect the instrument to the global cloud. The latter in tern opens the pathway to the effective data and code sharing, formation of the distributed human-ML teams, and emergence of the lateral instrumental networks.

However, the transition to the automated experiments also requires deep changes in the way scientific community plans and executes experimental activities. To date, all examples of the automated experiment in microscopy we are aware of are performed with the workflows based on fixed policies and a priori known objects of interest. The only examples of beyond human workflows include the inverse discovery experiments based on the deep kernel learning. Going beyond simple imitation of human operation and unleashing the power of automated experiment requires clearly defining the experimental reward, i.e. specific goals. This can include the discovery (curiosity learning), hypothesis falsification, or quantitative measurements. Many of these rewards are defined only within a broader scientific context of specific domain applications. Secondly, this requires formulating the deterministic or probabilistic policies, i.e. algorithms connecting the specific action expressed in the hyper language and the observed state of the system (image or spectra). These policies can be defined prior to the experiment to balance the exploration and exploitation goals. Alternatively, and much more interestingly, the policies can evolve along the experiment to achieve the desired reward within the given experimental budget.

Overall, the current state of the AE in STEM is nascent but fast changing. However, given the rapid emergence of the Python-based APIs and cloud infrastructure, remotely controlled microscopes, and especially given recent advances in active learning methods including Bayesian Optimization, reinforcement learning, and other forms of stochastic optimization, this field is likely to grow very rapidly in the coming years.

**Acknowledgements:**

This research (D.M., K.R., A.G., M.Z., A.A., N.S.R.), is sponsored by the INTERSECT Initiative as part of the Laboratory Directed Research and Development Program of Oak Ridge National Laboratory, managed by UT-Battelle, LLC, for the US Department of Energy under contract DE-AC05-00OR22725. S.V.K. was supported by the UT Knoxville start-up funding. This research used resources of the Center for Nanophase Materials Sciences, which is a DOE Office of Science User Facility. C.D., S.A., and S.R.S. were supported by the Chemical Dynamics Initiative / Investment, under the Laboratory Directed Research and Development (LDRD) Program at Pacific Northwest National Laboratory (PNNL). PNNL is a multi-program national laboratory operated for the U.S. Department of Energy (DOE) by Battelle Memorial Institute under Contract No. DE-AC05-76RL01830. J.C.A. acknowledges support from the Army/ARL via the Collaborative for Hierarchical Agile and Responsive Materials (CHARM) under cooperative agreement W911NF-19-2-0119, National Science Foundation under grant OAC:DMR:CSSI – 2246463, and US Department of Energy, Office of Science, Office of Advanced Scientific Computing Research under Award No. DE-SC-0002501. B.B. acknowledges support from NSF awards 1931306, 2226419, and 2209892. This work was supported in part by the U.S. Department of Commerce, National Institute of Standards and Technology as part of the Center for Hierarchical Materials Design (CHiMaD) under award number 70NANB19H005